\documentclass{ws-ijmpa}

\begin{document}

\markboth{Y. Ping, H. Liu and L. Xu} {Using (4+1) Split and Energy
Conditions to Study the Induced Matter in $5D$ Ricci-Flat
Cosmology}

%
\catchline{}{}{}{}{}
%

\title{USING (4+1) SPLIT AND ENERGY CONDITIONS TO STUDY THE INDUCED MATTER IN $5D$ RICCI-FLAT COSMOLOGY}

\author{YONGLI PING, HONGYA LIU\footnote{Corresponding author.} and LIXIN XU}

\address{School of Physics and Optoelectronic
Engineering, Dalian University of
Technology, \\
Dalian, Liaoning 116024, P.R.China\\
hyliu@dlut.edu.cn}

\maketitle

\begin{history}
\received{Day Month Year} \revised{Day Month Year}
\end{history}

\begin{abstract}
We use $(4+1)$ split to derive the $4D$ induced energy density
$\rho $ \ and pressure $p$ of the $5D$ Ricci-flat cosmological
solutions which are characterized by having a bounce instead of a
bang. The solutions contain two arbitrary functions of time $t$
and, therefore, are mathematically rich in giving various
cosmological models. By using four known energy conditions (null,
weak, strong, and dominant) to pick out and study physically
meaningful solutions, we find that the $4D$ part of the $5D$
solutions asymptotically approach to the standard $4D$ FRW models
and the expansion of the universe is decelerating for normal
induced matter for which all the four energy conditions are
satisfied. We also find that quintessence might be normal or
abnormal, depending on the parameter $w$ of the equation of state.
If $-1\leqslant w<-1/3$, the expansion of the universe is
accelerating and the quintessence is abnormal because the strong
energy condition is violated while other three are satisfied. For
phantom, all the four energy conditions are violated. Before the
bounce all the four energy conditions are violated, implying that
the cosmic matter before the bounce could be explained as a
phantom which has a large negative pressure and makes the universe
bouncing. In the early times after the bounce, the dominant energy
condition is violated while the other three are satisfied, and so
the cosmic matter could be explained as a super-luminal acoustic
matter.
\end{abstract}

\keywords{Kaluza-Klein theory, Cosmology, Energy conditions.}
\ccode{PACS numbers: 04.50.+h, 98.80.-k, 02.40.-k}

\section{Introduction}

The Campbell-Magaard theorem\cite{Campbell}\cdash\cite{Magaard}
states that any analytic solution of Einstein's equations in
$N$-dimensions can be locally embedded in a $(N+1)$-dimensional
Ricci-flat manifold. It was
noted\cite{Ripple}\cdash\cite{Seahra03} that this theorem provides
a mathematical support to the $5$-dimensional Space-Time-Matter
(STM) theory\cite{Wesson99}\cdash\cite{O+W} in which the $5D$
manifold is Ricci-flat with $R_{ab}=0$ while the $4D$ matter is
induced from the $5D$ vacuum. Many solutions of this type have
been studied within the framework of the STM theory. Here, in this
paper, we are going to study a class of $5D$ Ricci-flat
cosmological solutions which was firstly presented by Liu and
Mashhoon\cite{L+M95} and restudied latter by Liu and
Wesson\cite{L+W01}. This class of solutions is algebraically rich
because it contains two arbitrary functions of the time $t$. It
was shown\cite{L+W01} that two major properties characterize these
$5D$ models: Firstly, the $4D$ induced matter could be described
by a perfect fluid plus a variable cosmological term, and by
properly choosing the two arbitrary functions both the radiation
and matter dominated standard FRW models could be recovered.
Secondly, the big bang singularity of the $4D$ standard cosmology
could be replaced by a big bounce at which the size of the
universe is finite and the universe contracts before the bounce
and expands after the bounce. Many further studies could be found
in literature such as
those about the embeddings of the $5D$ solutions to brane models\cite%
{Seahra}\cdash\cite{Liu-plb}, about the accelerating expansion and
dark energy of the $5D$ universe\cite{XuWang}\cdash\cite{HYLiu}, and
about the isometry between the $5D$ solutions and $5D$ black
holes\cite{SearhaW}. We aware that it is the two arbitrary functions
that makes the $5D$ solutions mathematically rich in giving various
cosmological models and various normal and non-normal induced
matters. Therefore, it is useful to look for physical constraints to
fix these two arbitrary functions and to study physical properties
of various different solutions. This paper is organized as follows:
In Section II, we will use the $(4+1)$ split technique to derive the
$4D$ induced matter. In Section III, we will use four known energy
conditions (strong, weak, null, and dominant) to study properties of
the $4D$ induced matter. Section IV is a short conclusion.

\section{$(4+1)$ split and the induced matter}

The $5D$ Ricci-flat cosmological solutions read
\begin{equation}
dS^{2}=B^{2}dt^{2}-A^{2}\left(
\frac{dr^{2}}{1-kr^{2}}+r^{2}d\Omega ^{2}\right) -dy^{2},
\label{line element}
\end{equation}%
\begin{equation}
A^{2}=\left( \mu ^{2}+k\right) y^{2}+2{\nu }y+\frac{\nu
^{2}+C}{\mu ^{2}+k} ,  \label{A}
\end{equation}%
\begin{equation}
B=\frac{1}{\mu }\frac{\partial A}{\partial t}\equiv
\frac{\dot{A}}{\mu },  \label{B}
\end{equation}%
where $d\Omega ^{2}=d\theta ^{2}+\sin^2 \theta d\psi ^{2}$, $\mu
=\mu \left( t\right) $ and $\nu =\nu \left( t\right) $ are two
arbitrary functions, $k$ is the 3D curvature index ($k=\pm 1,0$),
$C$ is a constant. Because the $5D$ manifold (\ref{line
element})-(\ref{B}) is Ricci-flat, we have $I_{1}\equiv
R=0$, $I_{2}\equiv R^{ab}R_{ab}=0$, and%
\begin{equation}
I_{3}\equiv R^{abcd}R_{abcd}=\frac{72C^{2}}{A^{8}} ,  \label{I-3}
\end{equation}%
so $C$ is related to the $5D$ curvature.

The general $(4+1)$ split in STM theory has been given in Ref.~\refcite{Sajko98}%
, where it was shown that the $15$ $5D$ equations $R_{ab}=0$ decompose into $%
10+4+1$ $4D$ equations: ten for the $4D$ Ricci tensor, four for the
Gauss-Codazzi equations, and one for the scalar field. For our $5D$ metric (%
\ref{line element}) we see that the $4D$ hypersurfaces $\Sigma _{y}$ are $%
y\equiv $ constant and the normal vectors to $\Sigma _{y}$ are%
\begin{equation}
n^{a}=(0,0,0,0,1),\qquad n_{a}=(0,0,0,0,-1) .  \label{n-a}
\end{equation}%
So the projection tensor is
\begin{equation}
h_{ab}=g_{ab}+n_{a}n_{b} .  \label{h-ab}
\end{equation}%
The extrinsic curvature tensor $K_{ab}$, which describes the rate of changes
of $\Sigma _{y}$ as it moves in the normal direction, is defined by
\begin{equation}
K_{ab}=-h_{a}^{c}\nabla _{c}n_{b},  \label{Kab}
\end{equation}%
where $\nabla _{c}$ is the $5D$ covariant derivative operator. Meanwhile,
the lapse function $\Phi =1$ and the shift vector $N^{\alpha }=0$. These
greatly simplifies the calculations for $K_{ab}$. Using (\ref{n-a}) and (\ref%
{h-ab}) we find%
\begin{equation}
K_{ab}=\left(
\begin{array}{cc}
K\alpha \beta  & 0 \\
0 & 0%
\end{array}%
\right),   \label{Kab2}
\end{equation}%
with
\begin{equation}
K_{\alpha \beta }=-\frac{1}{2}\partial _{y}g_{\alpha \beta } .
\label{Kab3}
\end{equation}%
The $4D$ Ricci tensor $^{\left( 4\right) }R_{\alpha \beta }$ on the $4D$
hypersurface $\Sigma _{y}$ can be expressed in terms of $K_{\alpha \beta }$
by
\begin{equation}
^{\left( 4\right) }R_{\alpha \beta }=\partial _{y}K_{\alpha \beta
}-KK_{\alpha \beta }+2K_{\alpha \gamma }K_{\beta }^{\gamma },
\label{4DRicci}
\end{equation}%
where $K=g^{\alpha \beta }K_{\alpha \beta }$. Substituting the exact $5D$
solutions (\ref{line element}) - (\ref{B}) into (\ref{Kab3}) to calculate $%
K_{\alpha \beta }$ and then using $K_{\alpha \beta }$ in (\ref{4DRicci}), we
obtain the non-vanishing components of $^{\left( 4\right) }R_{\alpha \beta }$
being
\begin{eqnarray}
^{\left( 4\right) }R_{0}^{0} &=&-\frac{3\mu \dot{\mu}}{A\dot{A}} ,
\notag \\
^{\left( 4\right) }R_{1}^{1} &=&^{\left( 4\right)
}R_{2}^{2}=^{\left( 4\right) }R_{3}^{3}=-\left[ \frac{\mu
\dot{\mu}}{A\dot{A}}+\frac{2\left( \mu ^{2}+k\right)
}{A^{2}}\right]  .  \label{R0123}
\end{eqnarray}%
So the $4D$ Einstein tensor $^{(4)}G_{\beta }^{\alpha }\equiv \
^{(4)}R_{\beta }^{\alpha }-\delta _{\beta }^{\alpha }$ $^{\left( 4\right)
}R/2$ become
\begin{eqnarray}
^{(4)}G_{0}^{0} &=&\frac{3(\mu ^{2}+k)}{A^{2}} ,  \notag \\
^{(4)}G_{1}^{1} &=&^{(4)}G_{2}^{2}=^{(4)}G_{3}^{3}=\frac{\mu ^{2}+k}{A^{2}}+%
\frac{2\mu \dot{\mu}}{A\dot{A}} .  \label{G0123}
\end{eqnarray}%
According to Einstein' theory of general relativity, this $^{(4)}G_{\beta
}^{\alpha }$ implies an effective or induced $4D$ energy-momentum tensor $%
^{(4)}T_{\beta }^{\alpha }$ . Let $^{(4)}G_{\beta }^{\alpha }=^{(4)}T_{\beta
}^{\alpha }$ and suppose $^{(4)}T_{\beta }^{\alpha }$ being described by a
perfect fluid with density $\rho $ and pressure $p$,
\begin{equation}
^{(4)}T_{\alpha \beta }=(\rho +p)u_{\alpha }u_{\beta }-pg_{\alpha
\beta } ,  \label{T0-3}
\end{equation}%
where $u^{\alpha }$ is the $4$-velocity with $u^{\alpha }=(u^{0},0,0,0)$ and
$u^{0}u_{0}=1$, then we find
\begin{eqnarray}
\rho  &=&\frac{3(\mu ^{2}+k)}{A^{2}}\quad ,  \notag \\
p &=&-\frac{\mu ^{2}+k}{A^{2}}-\frac{2\mu \dot{\mu}}{A\dot{A}} .
\label{rho,p}
\end{eqnarray}%
Thus we rederived the same results for the $4D$ induced matter as
those in Ref.~\refcite{L+W01}. From the $5D$ solutions (\ref{line
element}) - (\ref{B}) one can see that if the two arbitrary
functions $\mu (t)$ and $\nu (t)$ are specified, the two scale
factors $A(t,y)$ and $B(t,y)$ and then the evolution of the
universe are fixed immediately. Then the cosmic induced matter
$\rho $\ and $p$ can be calculated with use of (\ref{rho,p}).
Meanwhile, the $4D$ Bianchi identities $^{(4)}G_{\beta ;\alpha
}^{\alpha }=0$ lead to the $4D$ conservation laws $^{(4)}T_{\beta
;\alpha }^{\alpha }=0$
which give%
\begin{equation}
\overset{\cdot }{\rho }=-3\frac{\overset{\cdot }{A}}{A}(\rho +p),
\label{conserv}
\end{equation}%
as is in the standard $4D$ cosmology.

\section{Energy conditions}

The energy conditions of Einstein's general relativity are designed to
extract as much information as possible from the field equations. They were
used in deriving many theorems such as the singularity theorems\cite%
{Howking}, the censorship theorem\cite{Censorship} and so on. In
this section, we wish to use the energy conditions to study the
properties of the induced matter of the $5D$ Ricci-flat
cosmological solutions (\ref{line element}) - (\ref{B}).

The standard classical energy conditions are the null energy
condition (NEC), weak energy condition (WEC), strong energy
condition (SEC), and dominant energy condition (DEC). Basic
definitions of these energy conditions can be found in
Ref.~\refcite{Wald}. For the case in cosmology they
are\cite{Visser97}\cdash\cite{Visser97-2}
\begin{eqnarray}
NEC &:&\qquad \rho +p\geq 0,  \label{nec} \\
WEC &:&\qquad \rho \geq 0\quad \text{and}\quad \rho +p\geq 0,  \label{wec} \\
SEC &:&\qquad \rho +3p\geq 0\quad \text{and}\quad \rho +p\geq 0,  \label{sec}
\\
DEC &:&\qquad \rho \geq 0\quad \text{and}\quad \rho \geqslant \left\vert
p\right\vert .  \label{dec}
\end{eqnarray}%
Here, for simplicity, we are not going to discuss models which contain more
than one components of matter. We just consider the case where the universe
is dominated by just one fluid with $\rho $ and $p$.

Without loss of generality, we write the equation of state as%
\begin{equation}
p=w\rho  .  \label{EOS}
\end{equation}%
Because both $\rho $\ and $p$ are functions of $t$ and $y$ in the $5D$
model, So $w$ is, in general, a function of $t$ and $y$ as well, $w=w(t,y)$.
Substituting (\ref{EOS}) into the four energy conditions (\ref{nec}) - (\ref%
{dec}), we get%
\begin{eqnarray}
NEC &:&\qquad (1+w)\rho \geq 0,  \label{nec-w} \\
WEC &:&\qquad \rho \geq 0\quad \text{and}\quad w\geq -1,  \label{wec-w} \\
SEC &:&\qquad (1+3w)\rho \geq 0\quad \text{and}\quad (1+w)\rho \geq 0,
\label{sec-w} \\
DEC &:&\qquad \rho \geq 0\quad \text{and}\quad \left\vert w\right\vert
\leqslant 1.  \label{dec-w}
\end{eqnarray}

Following previous usage (see, for example,
Ref.~\refcite{Visser97}-\refcite{Visser97-2}) we call matter that
satisfies all the four energy conditions ``normal'' and call matter
that specifically violates the SEC ``abnormal''. And we call matter
that violates any one of the four energy conditions ``non-normal''.
In the following we will discuss both the normal and non-normal
matter separately.

\subsection{Models dominated by normal matter}

For normal matter all the four standard energy conditions should be
satisfied. We can easily show that (\ref{nec-w}) - (\ref{dec-w}) are
equivalent to%
\begin{equation}
\text{Normal Matter}:\qquad \rho \geq 0\quad \text{and}\quad -1/3\leqslant
w\leqslant 1.  \label{NM}
\end{equation}%
With use of these constraints in the $5D$ induced matter (\ref{rho,p}) we
obtain%
\begin{eqnarray}
\mu ^{2}+k &\geqslant &0,  \label{mu-1} \\
-\frac{2\mu \dot{\mu}}{A\dot{A}} &\geqslant &0 ,  \label{mu-2} \\
\frac{2(\mu ^{2}+k)}{A^{2}}+\frac{\mu \dot{\mu}}{A\dot{A}}
&\geqslant &0 .  \label{mu-3}
\end{eqnarray}%
Meanwhile, with use of the definition of the proper time, $d\tau \equiv Bdt$%
, and the relation $B=\dot{A}/\mu $, the Hubble and deceleration parameters
of the $5D$ solutions are found to be%
\begin{equation}
H=\frac{\mu }{A},\qquad q\equiv -\frac{A\dot{\mu}}{\mu \dot{A}}.  \label{H,q}
\end{equation}%
Using (\ref{H,q}) in (\ref{mu-1}) - (\ref{mu-3}) we get%
\begin{eqnarray}
H^{2}+\frac{k}{A^{2}} &\geqslant &0 ,  \label{Hq-1} \\
q &\geqslant &0 ,  \label{Hq-2} \\
(2-q)H^{2}+\frac{2k}{A^{2}} &\geqslant &0 .  \label{Hq-3}
\end{eqnarray}%
We find that all the results in (\ref{NM}) and (\ref{Hq-1}) - (\ref{Hq-3})
are completely the same as those in $4D$ standard FWR cosmology for models
dominated by normal matter. And we also arrive, from (\ref{Hq-2}), at a
conclusion that \textit{the expansion of the universe is decelerating if the
universe is dominated by normal matter, and so the gravitational force of
normal matter is attractive}. This conclusion valid for both the $4D$
standard FRW cosmology and the $5D$ Ricci-flat cosmology.

As we mentioned above that $w$ in (\ref{EOS}) is in general not a constant.
Now we assume $w$ being a constant for simplicity. Then Eqs. (\ref{rho,p})
lead to%
\begin{equation}
\frac{\left( 3w+1\right) \dot{A}}{A}=-\frac{2\mu \dot{\mu}}{\mu
^{2}+k} .  \label{mu,A1}
\end{equation}%
Integrating with respect to $t$, we obtain%
\begin{equation}
\mu ^{2}+k=\frac{c_{1}}{A^{3w+1}}\qquad \text{for \ }w=\text{constant }.
\label{mu2+k}
\end{equation}%
And the conservation law(\ref{conserv}) gives%
\begin{equation}
\rho =\rho _{0}\left( \frac{A_{0}}{A}\right) ^{3(1+w)}\qquad \text{for \ }w=%
\text{constant }.  \label{rho-A}
\end{equation}%
Let us return to the two arbitrary functions\ $\mu (t)$ and $\nu (t)$ of the
$5D$ solutions. From the $5D$ solution (\ref{A}) we can see that the second
arbitrary function $\nu (t)$\ can always be solved out in terms of $A(t)$ if
we substituting (\ref{mu2+k}) in (\ref{A}). This leaves us a freedom to
consider the equation (\ref{mu2+k}) alone. Recall that the Campbell-Magaard
theorem just says that the $4D$ solutions of Einstein equations can be
embedded in a $5D$ Ricci-flat manifold \textit{locally (}not\textit{\
globally)}. This reminds us to choose the time $t$ in such a way that the
coordinate time $t$ approaches to the proper time asymptotically for a very
large $t$. Consider the $k=0$ case for simplicity. Then, to make $B=\dot{A}%
/\mu \rightarrow 1$, we must have, from (\ref{mu2+k}),%
\begin{eqnarray}
A &\approx &A_{0}\left( \frac{t}{t_{0}}\right) ^{n},\qquad \text{ }\mu
\approx \frac{n}{t_{0}}A_{0}\left( \frac{t}{t_{0}}\right) ^{n-1},  \notag \\
\qquad n &=&\frac{2}{3(w+1)},\qquad \text{for }t\gg 0 .
\label{mu-t}
\end{eqnarray}%
Then we get an approximate $5D$ metric%
\begin{eqnarray}
dS^{2} &\approx &dt^{2}-A_{0}^{2}\left( \frac{t}{t_{0}}\right) ^{2n}\left(
dr^{2}+r^{2}d\Omega ^{2}\right) -dy^{2},  \notag \\
n &=&\frac{2}{3(w+1)}\qquad \text{for }t\gg 0 .  \label{metricApp}
\end{eqnarray}%
This is a class of approximate solutions of the $5D$ Ricci-flat universe
while the corresponding exact solutions are given in (\ref{mu2+k}) and (\ref%
{rho-A}). Now we calculate the deceleration parameter $q$. We find that for
both the exact (with $k=0$) and approximate solutions, we get the same
result,%
\begin{equation}
q=\frac{3w+1}{2} .  \label{q-nonnor}
\end{equation}%
As we mentioned above that the constant $w$ takes values from $-1/3$ to $1$
for normal matter for which $q$ takes the values from $0$ to $2$. So the
expansion of the universe is decelerating for normal matter. We should also
point out that this approximate solution is actually valid for any values of
$w$ as we will discuss in the next subsection. Thus we see from (\ref%
{metricApp}) that the $4D$ part of the $5D$ universe asymptotically
approaches to that of the standard FRW cosmology in the case for a normal
induced matter.

\subsection{Models dominated by non-normal matter}

Let us consider the $5D$ approximate solution (\ref{mu-t}) - (\ref{metricApp}%
) and their corresponding exact solutions in (\ref{mu2+k}) and (\ref{rho-A})
where $w$ takes values from $-1/3$ to $1$ for normal matter. Now we discuss
the case for $w<-1/3$, for which at least one of the four energy conditions
must be violated and so the induced matter is non-normal.

From the four energy conditions (\ref{nec-w}) - (\ref{dec-w}) we can see
that if $-1\leqslant w<-1/3$, SEC is violated while NEC, WEC, and DEC are
satisfied. This kind of matter is usually called abnormal matter. We can
also see that if $w<-1$, all the four energy conditions, NEC, WEC, SEC, and
DEC, are violated. In what follows we will discuss these two cases
separately.

\subsubsection{Models with $-1\leqslant w<-1/3$}

For this case, Eq. (\ref{q-nonnor}) gives that $-1\leqslant q<0$. So
the expansion of the universe is accelerating. Thus we recover the
$5D$ cosmological scaling solution\cite{Chang} in which the $4D$
induced matter is of a quintessence model of scalar field. In the
quintessence dark energy model\cite{Zlatev}\cdash\cite{Steinhardt},
$w$ is in the range [$-1$, $1$]. So quintessence might be a normal
matter if $-1/3\leqslant w<1$ or abnormal matter if $-1\leqslant
w<-1/3$. This conclusion valid in both the $4D$ standard and $5D$
Ricci-flat cosmological models.

\subsubsection{Models with $w<-1$}

For this case we have $q<-1$. Note that from (\ref{mu-t}) we find that $n$
is negative and $A(t)\propto t^{n}$ is decreasing, contrary to the
assumption for an expanding universe. To resolve this problem, we find that,
without violating their correctness as an approximate solution, the forms of
(\ref{mu-t}) and (\ref{metricApp}) can be changed to%
\begin{eqnarray}
A &\approx &A_{0}\left( \frac{t_{m}-t_{0}}{t_{m}-t}\right) ^{n},\qquad \text{
}\mu \approx \frac{n}{t_{m}-t_{0}}A_{0}\left( \frac{t_{m}-t_{0}}{t_{m}-t}%
\right) ^{n-1},  \notag \\
n &=&-\frac{2}{3(w+1)},\qquad \text{for }0\ll t<t_{m} .
\label{mu-tPh}
\end{eqnarray}%
and%
\begin{eqnarray}
dS^{2} &\approx &dt^{2}-A_{0}^{2}\left( \frac{t_{m}-t_{0}}{t_{m}-t}\right)
^{2n}\left( dr^{2}+r^{2}d\Omega ^{2}\right) -dy^{2},  \notag \\
n &=&-\frac{2}{3(w+1)}\qquad \text{for }0\ll t<t_{m} .
\label{metricAppPh}
\end{eqnarray}%
Thus we recover the $5D$ attractor solution\cite{HYLiu} in which
the $4D$ induced matter is described by a phantom model of the
dark energy. As $t$ tends to $t_{m}$, the scale factor $A(t)$
tends to infinity implying that the universe will undergo a big
rip and will expand to infinity within a finite time. Here we see
that all the four energy conditions are violated for phantom,

\subsection{Before and during the bounce}

The $5D$ Ricci-flat cosmological models are characterized by having a big
bounce instead of a big bang. Now we wish to know what kind of non-normal
matter dominated the universe before and during the bounce. To make the
discussion easier, we list an exact solution\cite{XuWang} in the following.

\begin{eqnarray}
k &=&0,\ \ \ C=1,  \notag \\
\nu \left( t\right) &=&{t_{c}}/{t},\ \ \ \mu \left( t\right) =t^{{-1}/{2}},
\label{k,C}
\end{eqnarray}%
where $t_{c}$ is a constant. Substituting this equation into (\ref{A}), (\ref%
{B}) and (\ref{rho,p}), we obtain
\begin{equation}
A^{2}=t\left[ 1+\left( \frac{y+t_{c}}{t}\right) ^{2}\right],
\label{Abounce}
\end{equation}%
\begin{equation}
B^{2}=\frac{1}{4}\left[ 1-\left( \frac{y+t_{c}}{t}\right) ^{2}\right] ^{2}%
\left[ 1+\left( \frac{y+t_{c}}{t}\right) ^{2}\right] ^{-1},  \label{Bbounce}
\end{equation}%
and
\begin{equation}
\rho =\frac{3}{t^{2}\left[ 1+\left( (y+t_{c})/t\right) ^{2}\right] },
\label{rhobounce}
\end{equation}%
\begin{equation}
p=\frac{2}{t^{2}\left[ 1-\left( (y+t_{c})/t\right) ^{2}\right] }-\frac{1}{%
t^{2}\left[ 1+\left( (y+t_{c})/t\right) ^{2}\right] }.  \label{pbounce}
\end{equation}

From Eq.(\ref{Abounce}) we can show that the scalar factor $A\left(
t,y\right) $ has a minimum point at
\begin{equation}
t=\mid {y+t_{c}}\mid \equiv {t_{b}}.  \label{t-b}
\end{equation}%
at which we have
\begin{equation}
A\mid _{t=t_{b}}=(2t_{b})^{1/2},\ B\mid _{t=t_{b}}=0,\ \dot{A}\mid
_{t=t_{b}}=0.  \label{A-tb}
\end{equation}%
Here $t=t_{b}$ is the big bounce singularity.

In this solution, the time $t$ is defined in the range $(0$, $\infty )$.
Generally, $t$ is not the proper time. However, when $t\rightarrow \infty $,
we have $B\rightarrow 1/4$ and $A\rightarrow t^{1/2}$. so the coordinate
time $t$ tends to the proper time for very large $t$. The bounce is at $%
t=t_{b}$. Before the bounce means for $0<t<t_{b}$. When $t\rightarrow 0$, we
have $A\rightarrow \infty $ and $B\rightarrow 1/(2t)$. When changed to the
proper time $\tau $, we find $t\propto e^{2\tau }$. So $t\rightarrow 0$
corresponds to $\tau \rightarrow -\infty $ and $A\rightarrow \infty $.

From this solution we can show that before the bounce (for $0<t<t_{b}$ ) we
have%
\begin{eqnarray}
\rho  &>&0,\quad p<0,  \label{rho+3p} \\
\rho +p &<&0,\quad \rho +3p<0.  \notag
\end{eqnarray}%
We find that before the bounce all the four energy conditions, NEC, WEC,
SEC, and DEC, are violated. Meanwhile, because $\rho >0$ and $p<-\rho $, so $%
w<-1$. So, before the bounce, the non-normal induced matter has a large
negative value and generates a repulsive force. Then the bounce could be
explained as due to this repulsive force and the induced matter could be
explained as a phantom.

We can also shown that for $t_{b}<t<\sqrt{3}t_{b}$, we have $\rho >0$, $p>0$%
, $\rho +p>0$, $\rho +3p>0$, and $\rho -|p|<0$. So NEC, WEC and SEC are
satisfied while DEC is violated. During this period, the induced cosmic
matter is perhaps a super-luminal acoustic matter.

\section{Conclusion}

In this paper we have used the $(4+1)$ split to derive the $4D$
induced energy-momentum tensor of the $5D$ Ricci-flat cosmological
solutions. Then we have used the four energy conditions, NEC, WEC,
SEC, and DEC, to discuss the physical properties of the induced
matter and the various solutions of the $5D$ models. Firstly, we
have shown that if the universe is dominated by normal matter, the
expansion of the universe is decelerating and the $4D$ part of the
$5D$ Ricci-flat universe approaches asymptotically to the $4D$
standard FRW models in late times of the universe. If the universe
is dominated by quintessence, the induced matter could be normal
or abnormal, depending on the value of $w$. If $-1/3\leqslant
w<1$, all the four energy
conditions are satisfied and the quintessence is a normal matter. If $%
-1\leqslant w<-1/3$, only the SEC is violated and so in this range the
quintessence is abnormal and the expansion of the universe is accelerating.
For phantom, all the four energy conditions are violated. We also find that
before the bounce all the four energy conditions are violated, and so the
induced matter could be explained as phantom and the bouncing could be
explained as due to the repulsive force of the phantom. In the early time
after the bounce, DEC is violated while the other three are satisfied, so
the induced cosmic matter in this period could be explained as a
super-luminal acoustic matter.

\section*{Acknowledgments}

We would like to thank NEF (10573003) and NBRP (2003CB716300) of
P.R. China for financial support.

\end{document}